\documentclass[
5p,
preprint,
]{elsarticle}

\usepackage[%
  colorlinks=true,
  urlcolor=blue,
  linkcolor=blue,
  citecolor=blue
]{hyperref}
\usepackage{amsmath}
\usepackage{amssymb}
\usepackage[all]{hypcap}
\usepackage{graphicx}
\usepackage{dcolumn}
\usepackage{bm}
\usepackage{booktabs}

\usepackage[utf8]{inputenc}
\usepackage[T1]{fontenc}
\usepackage{mathptmx}

\begin{document}
\begin{frontmatter}

\title{Influence of Molecular Beam Effusion Cell Quality on Optical and Electrical Properties of Quantum Dots and Quantum Wells}

\author[mymainaddress,mythirdaddress]{G. N. Nguyen \corref{mycorrespondingauthor}}
\cortext[mycorrespondingauthor]{Corresponding author at: Lehrstuhl für Angewandte Festkörperphysik, Ruhr-Universität Bochum, Germany}
\ead{giang.nguyen@alumni.ruhr-uni-bochum.de}

\author[mymainaddress]{A. R. Korsch}
\author[mymainaddress]{M. Schmidt}
\author[mymainaddress]{C. Ebler}
\author[mymainaddress]{P. A. Labud}
\author[mymainaddress]{R. Schott}
\author[mymainaddress,mysecondaryaddress]{P. Lochner}
\author[mymainaddress,mysecondaryaddress]{F. Brinks}
\author[mymainaddress]{A. D. Wieck}
\author[mymainaddress]{and A. Ludwig}

\address[mymainaddress]{Lehrstuhl f\"ur Angewandte Festk\"orperphysik, Ruhr-Universit\"at Bochum, Universit\"atsstra\ss e 150, 44780, Germany.}
\address[mysecondaryaddress]{Fakult\"at f\"ur Physik and CENIDE, Universit\"at Duisburg-Essen, Lotharstra\ss e 1, 46048 Duisburg, Germany.}
\address[mythirdaddress]{Department of Physics, University of Basel, Klingelbergstrasse 82, 4056 Basel, Switzerland.}

\date{\today}

\begin{abstract}
Quantum dot heterostructures with excellent low-noise properties became possible with high purity materials recently. We present a study on molecular beam epitaxy grown quantum wells and quantum dots with a contaminated aluminum evaporation cell, which introduced a high amount of impurities, perceivable in anomalies in optical and electrical measurements. We describe a way of addressing this problem and find that reconditioning the aluminum cell by overheating can lead to a full recovery of the anomalies in photoluminescence and capacitance-voltage measurements, leading to excellent low noise heterostructures. Furthermore, we propose a method to sense photo-induced trap charges using capacitance-voltage spectroscopy on self-assembled quantum dots. Excitation energy-dependent ionization of defect centers leads to shifts in capacitance-voltage spectra which can be used to determine the charge density of photo-induced trap charges via 1D band structure simulations. This method can be performed on frequently used quantum dot diode structures.
\end{abstract}


\end{frontmatter}

\section{\label{Introduction} Introduction}

Semiconductors are indispensable in a large variety of applications.
They act as reliable switches in low and high power applications \cite{j.s.h.schoenberg_Ultrawideband_1997}, efficient computational units \cite{loss_quantum_1998} like central processing units, energy conversion devices and optoelectronic devices like photodetectors, LEDs and lasers \cite{shields_semiconductor_2007}. 
The success of semiconductors and their dominance over other material devices originates from their variety of parameters, like small to wide band gaps and the ability to combine them, leading to controllable quantum effects in nanostructures due to small effective carrier masses. 
This success always went along with a struggle to achieve appropriate material and device quality.
For optoelectronic devices, in particular, special care has to be taken to minimize non-radiative recombination centers.
Arsenide based semiconductors are a mature material system to be found in many electronic applications. They are important for a huge variety of fundamental physics discoveries and handled as candidates for future quantum devices \cite{warburton_single_2013}, for which new and much higher demands to material quality are needed \cite{smith_Lowdimensional_1996, wang_gallium_2014}.
Reduced dimensions by constrictions on the nanometer scale in spatially tailored semiconductors lead to quantum confinement and large local density of states in a desired energy range, which is a huge advantage for many optoelectronic applications.
One platform of such carrier confinement are self-assembled quantum dots (QDs), which enable research in fundamental solid-state physics \cite{lobl_radiative_2020,korsch_Temperature_2019} and quantum communication due to their excellent optical properties \cite{zhai_LowNoise_2020, liu_solidstate_2019a}.
While the disadvantage of non-radiative recombination centers is obvious for optoelectronic devices, fluctuating charges in the environment of QDs are also unfavorable. Imperfections, like trap states caused by defects, impurities or interfaces lead to spectral wandering or even blinking, as the electric field varies randomly and changes the emitter energy levels by the quantum-confined Stark effect \cite{alen_Starkshift_2003}. Furthermore, even the charge state of the emitter changes due to trap states \cite{houel_probing_2012}.
This is not desirable for envisaged applications of quantum systems like qubits or the generation of indistinguishable single photons. For this reason, finding methods to reduce imperfections is of essential importance. The QD energy level is a very sensitive probe for charge fluctuations \cite{vamivakas_nanoscale_2011a, kuhlmann_charge_2013}. In the work by Houel et al., the authors use the transition energy from an excited state to the ground state of a single QD to detect and localize single charge quanta \cite{houel_probing_2012}. Another method to detect fluctuations was reported by Mooney who observed a shift of capacitance-voltage (CV) spectra due to photo-induced ionization of deep donor levels in the AlGaAs region of a MODFET \cite{mooney_deep_1990}.\\
\indent Using the methods shown in this paper, heterostructures with low level of these impurities can be produced, enabling quantum optics-, spin and nuclear spin experiments \cite{prechtel_Decoupling_2016b,allison_Tuning_2014,kuhlmann_Transformlimited_2015,javadi_Spin_2018,najer_gated_2019,lobl_Excitons_2019a}. In the following, we show the influence of the usage of impure aluminum or a contaminated aluminum evaporation cell in MBE on optical and electrical properties of heterostructures. We present a method to enhance the optical properties by reconditioning the evaporation cell by overheating. Additionally, we present an approach using the shift of peak positions in the CV charging spectrum of the QD ensemble to detect photo-induced activation of defect centers. We find an excitation energy-dependent ionization of defect centers. Using 1D band structure simulation, we develop a model to quantify the illumination induced trap charge density, which is an indicator of the sample quality.
\section{\label{experiment} Sample and Growth description}
\begin{sloppypar}
\tolerance 9999
\begin{table*}[h]
\centering
\caption{Sample overview and growth parameters. The sample temperature was determined using a pyrometer. Note that Al-cell-1 is the contaminated Al-cell, Al-cell-2 is the intact Al-cell and Al-cell-1$^*$ is the reconditioned Al-cell-1. Numbers behind sample labels are for internal reference. }
\begin{tabular}{lc|lcc|l}
\toprule
Sample-          & Temperature & Al-Cell- &Temperature &Overheating Temperature $\Delta T$  & Comment \\
\midrule
$A_{\mathrm{PL,C2}}$ (\#14352) & 620\,$^\circ$\,C    & 2 & 1102\,$^\circ$\,C    & 98\,$^\circ$\,C           & good PL \\
$B_{\mathrm{PL,C1}}$ (\#14347) & 620\,$^\circ$\,C    & 1 & 1164\,$^\circ$\,C    &  36\,$^\circ$\,C         & bad PL  \\
$C_\mathrm{PL,C2}$ (\#14984) & 615\,$^\circ$\,C    & 2 & 1102\,$^\circ$\,C    & 128\,$^\circ$\,C            & good PL \\
$D_{\mathrm{PL,C1^*}}$ (\#14979) & 615\,$^\circ$\,C    & 1$^*$ & 1163\,$^\circ$\,C    & 67\,$^\circ$\,C            & bad PL \\
$E_\mathrm{PL,C1^*}$ (\#15014) & 615\,$^\circ$\,C    & $1^*$ & 1164\,$^\circ$\,C    & 100\,$^\circ$\,C (2\,h) + 115\,$^\circ$\,C (20\,min)           & good PL\\
\midrule
$A_\mathrm{CV,C1^*}$ (\#14729) & 600\,$^\circ$\,C    & 1$^{*}$  & 1177\,$^\circ$\,C    & 23\,$^\circ$\,C            & good CV  \\
$B_\mathrm{CV,C1}$ (\#14331) & 600\,$^\circ$\,C    & 1& 1164\,$^\circ$\,C    & 36\,$^\circ$\,C          & bad CV \\
\midrule
$A_\mathrm{Hall,C2}$ (\#14351) & 630\,$^\circ$\,C    & 2  & 1102\,$^\circ$\,C    & 98\,$^\circ$\,C           & good Hall  \\
$B_\mathrm{Hall,C1}$ (\#14330) & 630\,$^\circ$\,C    & 1& 1164\,$^\circ$\,C    &    36\,$^\circ$\,C       & good Hall \\
\bottomrule
\end{tabular}
\label{tab_samples}
\end{table*}
The samples under investigation are grown by a III-V-Riber-Epineat MBE in two growth campaigns but with identical gallium, arsenic and vacuum conditions. The aluminum, however, was deposited with two different pyrolytic boron nitride cold lip effusion cells of 60\,cm$^3$ (Al-cell-1) and 35\,cm$^3$ (Al-cell-2) volume. Samples grown with Al-cell-1 showed anomalies in CV and PL measurements (hysteresis in CV measurements and degradation of PL signal), which will be described in more detail in the following. Reference samples without anomalies in CV and PL measurements are grown with the backup Al-cell-2 and with the reconditioned Al-cell-1$^{*}$. \\
The samples are labelled as sample $A_{\mathrm{PL,C2}}$, $B_{\mathrm{PL,C1}}$, $C_{\mathrm{PL,C2}}$, $D_{\mathrm{PL,C1^*}}$, $E_{\mathrm{PL,C1^*}}$, $A_{\mathrm{CV,C1^*}}$, $B_{\mathrm{CV,C1}}$ and $A_\mathrm{Hall,C2}$, $B_\mathrm{Hall,C1}$ in the following: The samples $A_{\mathrm{PL,C2}}$ and $B_{\mathrm{PL,C1}}$ consist of  GaAs quantum wells (QW) of $16$\,nm, $12$\,nm, $10$\,nm, $8$\,nm, $7$\,nm, $6$\,nm, $5.5$\,nm, and $5$\,nm thickness with Al$_{0.34}$Ga$_{0.66}$As cladding/barrier material. Sample $A_{\mathrm{PL,C2}}$ was grown with the intact Al-cell-2, while sample $B_{\mathrm{PL,C1}}$ was grown using the contaminated Al-cell-1.
Samples $C_{\mathrm{PL,C2}}$, $D_{\mathrm{PL,C1^*}}$, and $E_{\mathrm{PL,C1^*}}$ consist of GaAs QWs of $16$\,nm, $10$\,nm, $7.5$\,nm, $6$\,nm, and $5$\,nm thickness with Al$_{0.34}$Ga$_{0.66}$As cladding/barrier material. Sample $C_{\mathrm{PL,C2}}$ is grown using Al-cell-2 whereas samples $D_{\mathrm{PL,C1^*}}$ and $E_{\mathrm{PL,C1^*}}$ are grown using Al-cell-1$^*$. \\
The samples $A_{\mathrm{CV,C1^*}}$ and $B_{\mathrm{CV,C1}}$ are bias-tuneable structures with self-assembled InAs QDs grown in the Stranski-Krastanov growth mode. The samples consist of a $300$\,nm back contact layer of n-type doped GaAs, which works as an electron reservoir, grown on a (100)-oriented GaAs substrate. A $25$\,nm GaAs tunnel barrier separates the QDs from the back contact. The QDs are capped by $11$\,nm of GaAs followed by a $154$\,nm blocking barrier of $\mathrm{Al}_{0.34}\mathrm{Ga}_{0.66}\mathrm{As}$ and a Schottky gate. While $B_{\mathrm{CV,C1}}$ was grown using Al-cell-1, $A_{\mathrm{CV,C1^*}}$ was grown with the reconditioned Al-cell-1$^*$ in another growth campaign after opening of the MBE. A more thorough sample growth description is given by Ludwig et al. \cite{ludwig_Ultralow_2017}.\\
In addition, mobility measurements on two single heterojunction delta-doped $50\,$nm setback high electron mobility transistor (HEMT) structures $A_\mathrm{Hall,C2}$ and $B_\mathrm{Hall,C1}$, grown with Al-cell-1 and Al-cell-2, are compared.\\
Table \ref{tab_samples} gives an overview of the characterized samples and the growth conditions. For each sample, an Al-cell overheating temperature $\Delta T$ is presented. It indicates the temperature difference between growth temperature and the maximum overheating cell temperature before the growth, which we will show to be an important factor in achieving high-quality samples.

\end{sloppypar}

\section{Experiment}
\begin{sloppypar}
\tolerance 9999
Optical spectroscopy is performed via PL measurements on samples $A_{\mathrm{PL,C2}}$, $B_{\mathrm{PL,C1}}$, $C_{\mathrm{PL,C2}}$, $D_{\mathrm{PL,C1^*}}$, and $E_{\mathrm{PL,C1^*}}$. The samples are excited above-band using a red $635$\,nm laser at a power of 3\,mW and focused onto the sample reaching an intensity of approximately 240\, W/cm$^2$. The emitted light is analyzed using a monochromator and the samples are cooled to $77$\,K using liquid nitrogen.\\
\indent Electrical measurements are performed using CV spectro\-scopy on samples $A_{\mathrm{CV,C1^*}}$ and $B_{\mathrm{CV,C1}}$. Through CV spectro\-scopy, it is possible to analyze the charging energies of QDs \cite{drexler_spectroscopy_1994, warburton_coulomb_1998}. This method records the differential capacitance of a diode with QDs embedded in the depletion zone, tunnel coupled to a reservoir of carriers. Varying the voltage applied to the sample leads to a tilting of the bands around the back contact, which works as the pivot point. We conduct CV spectroscopy by overlaying a small ac signal of $20$\,mV rms amplitude at $10834$\,Hz over the dc voltage $V_{\mathrm{G}}$ used to sweep through the QD energy levels. If a QD energy level is in resonance with the Fermi-level of the back contact, electron tunneling is observed with an increased current between gate and back contact which we measure with a Lock-In amplifier. This can be converted into a capacitance change between the gate and back contact. Sweeping $V_{\mathrm{G}}$ leads to a characteristic charging spectrum with peaks at voltages where electron tunneling into QDs is enhanced {(cf. Fig.~\ref{comparison_illu_noillu} and Fig.~\ref{gute_probe}}). The first two peaks are labeled $\mathrm{s}_1$ and $\mathrm{s}_2$ and are separated due to Coulomb repulsion followed by the charging of four p-peaks. The gate voltages are converted into energies with a simple lever approach \cite{warburton_coulomb_1998}. All measurements were conducted at $T~=~4.2$\,K inside a liquid He vessel and the samples were illuminated using commercial available LEDs. The measurement process is described in the following:\\
Throughout the cool-in process, all contacts and gates are short-circuited to prevent freezing of charges. At $T~=~4.2$\,K the sample is illuminated for 5\,min at a constant gate voltage $V_{\mathrm{illum}}$. Five different LEDs with following peak wavelength at $T~=~4.2$\,K were used: $920$\,nm, $860$\,nm, $600$\,nm, $510$\,nm, and $460$\,nm. Immediately after illumination, bidirectional CV measurements are performed, sweeping forward ($V_{\mathrm{G}}~=~-1.0$\,V to $+0.4$\,V) and backward ($V_{\mathrm{G}}~=~+0.4$\,V to $-1.0$\,V) in steps of $5$\,mV. This was repeated for each wavelength and for $V_{\mathrm{illum}}~=~-4.0$\,V to $2.0$\,V in steps of $0.2$\,V.
\end{sloppypar}

\section{\label{results} Results and Interpretation}

\subsection{\label{PL_section} Photoluminescence Characterization}
\begin{figure}[h!]
\centering
\includegraphics[width = 8.5cm]{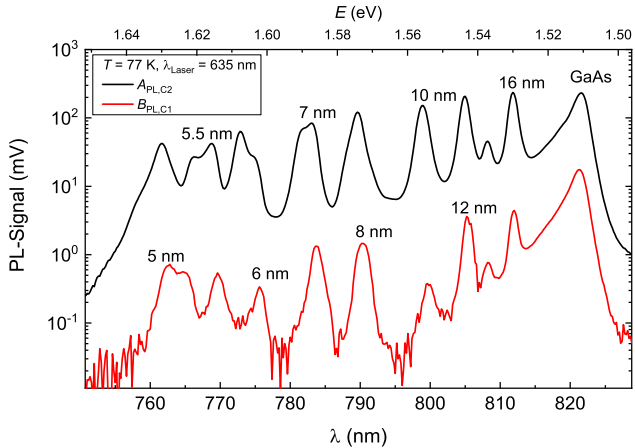}
\caption{Comparison of PL signal at $77$\,K of QWs of different sizes ($5\,\mathrm{nm}~\mathrm{to}~16\,\mathrm{nm}$) grown with Al-cell-1 (Sample $B_{\mathrm{PL,C1}}$, red) and QWs grown with Al-cell-2 (Sample $A_{\mathrm{PL,C2}}$, black). A difference in two orders of magnitude in signal amplitude is visible.}
\label{quantumwell}
\end{figure}

\begin{sloppypar}
\tolerance 9999
PL measurements on GaAs QWs grown with Al-cell-1 looked qualitatively comparable to that of a sample grown with the backup Al-cell-2. However, the quantum efficiency was orders of magnitude worse indicated by a significantly lower intensity. Fig. \ref{quantumwell} shows the PL signal at $T = 77$\,K of samples $A_{\mathrm{PL,C2}}$ and $B_{\mathrm{PL,C1}}$ consisting of a series of QWs of different sizes. Sample $B_{\mathrm{PL,C1}}$, grown with Al-cell-1 has a signal which is two orders of magnitude worse than for sample $A_{\mathrm{PL,C2}}$ grown with Al-cell-2.
\end{sloppypar}
\begin{figure}[h]
\centering
\includegraphics[width = 8.5cm]{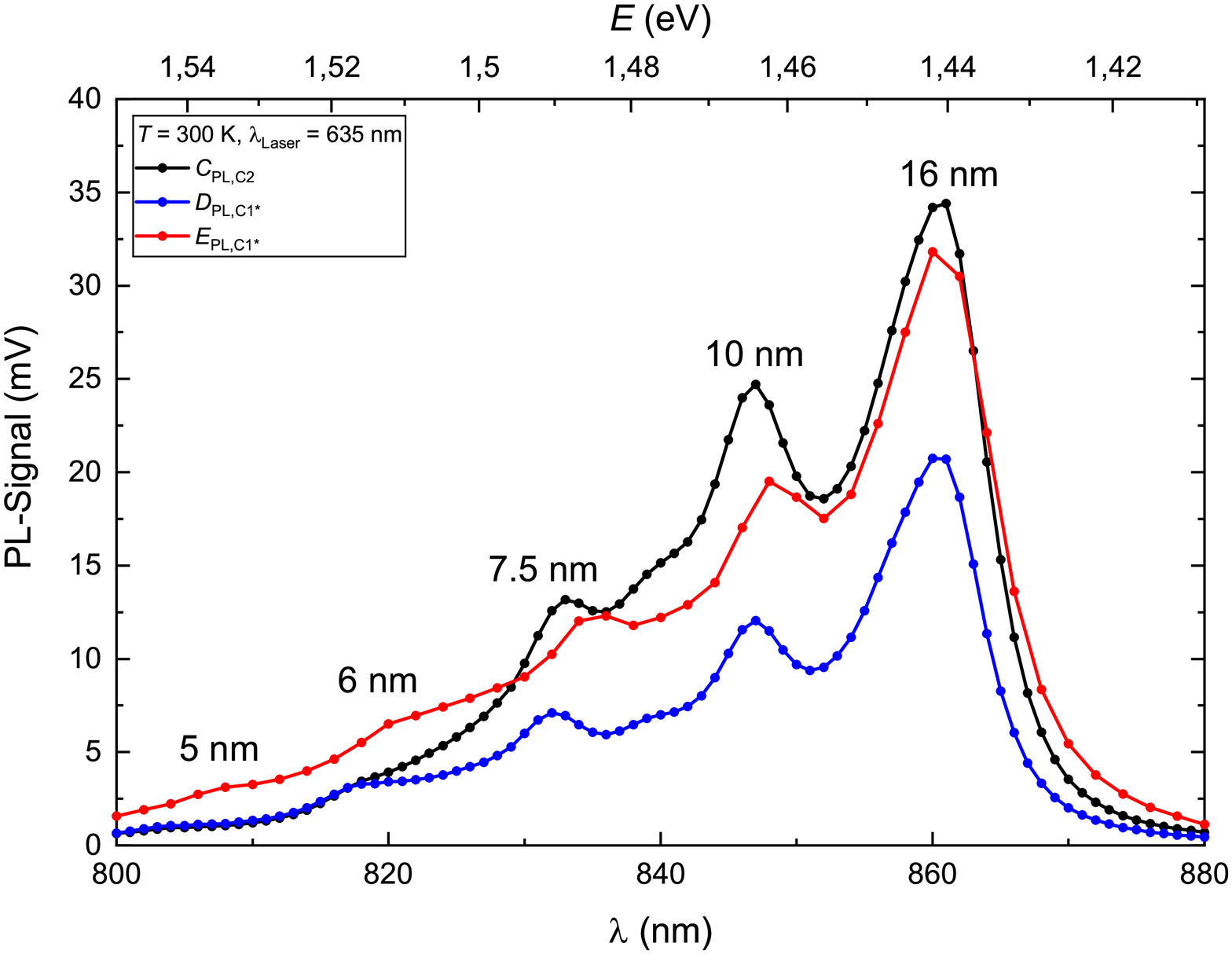}
\caption{Room temperature PL signal of samples $C_{\mathrm{PL,C2}}$ (black), $D_{\mathrm{PL,C1^*}}$ (blue), and $E_{\mathrm{PL,C1^*}}$ (red). After overheating the Al-cell with $100^{\circ}$\,C for 2h and $115^{\circ}$\,C for 20 min before the growth of sample $E_{\mathrm{PL,C1^*}}$ a similar PL intensity compared to sample $C_{\mathrm{PL,C2}}$ (black) grown with Al-cell-2 is achieved. }
\label{PL}
\end{figure}
\begin{sloppypar}
\tolerance 9999
The decrease of PL intensity on QWs could be recovered after reconditioning Al-cell-1 by overheating. With this, a signal comparable to the signal of samples grown using Al-cell-2 could be achieved. 
Fig. \ref{PL} shows the PL signal of samples $C_{\mathrm{PL,C2}}$, $D_{\mathrm{PL,C1^*}}$, and $E_{\mathrm{PL,C1^*}}$. 
Sample $C_{\mathrm{PL,C2}}$ is grown with Al-cell-2, whereas the other two are grown with the contaminated Al-cell-1 after partial recovery of the cell by reconditioned overheating. Typical conditioning of the cell before each growth day includes a cell ramp to $1200^{\circ}$\,C for $10$\,min while the effusion cell temperature for standard growth parameters (growth rate of $0.1\,\mathrm{nm/s}$) is $T=1164^{\circ}$\,C. A first overheating iteration was performed before the growth of sample $D_{\mathrm{PL,C1^*}}$ as other samples grown using Al-cell-1 did not show any signal in PL. During the first iteration, Al-cell-1 and Al-cell-2 were ramped to $1230^{\circ}$\,C with open shutters and overheated for $3$\,h. The shutters were opened and shut in $30$\,min intervals to also overheat the shutter plates. Sample $D_{\mathrm{PL,C1^*}}$ still showed a lower PL signal than the reference sample $C_{\mathrm{PL,C2}}$ grown using Al-cell-2. In a second iteration, the Al-cell was ramped to $1264^{\circ}$\,C and overheated for 2\,h and further ramped to $1279^{\circ}$\,C and overheated for another 20\,min. Here, the shutters were kept open the whole time. Sample $E_{\mathrm{PL,C1^*}}$, grown after the second iteration of overheating, showed a PL-signal comparable to the reference sample $C_{\mathrm{PL,C2}}$. Note that after recovering the PL-signal, no further overheating iterations were needed to maintain a good PL-signal. On the other hand, PL measurements on InAs QDs (cf. Ludwig et al. \cite{ludwig_Ultralow_2017} for growth conditions) grown with Al-cell-1 did not significantly deviate from samples grown with the reference cell Al-cell-2 (not shown here). Nevertheless, all anomalies that are presented in the following vanished after reconditioning of the contaminated Al-cell-1 reinforcing the assumption of impurities induced by Al-cell-1.\\
We explain the decrease in PL-signal for QWs with non-radiative recombination centers induced by impurities and defects during growth in the QW barrier. As PL experiments are performed using above-band excitation, charge carriers are captured in the impurities in the barrier leading to a decrease in PL signal. By reconditioning the cell by overheating, the PL signal can be enhanced close to the signal from the reference sample. This can be understood as a purification of the Al-cell as with increased cell temperature more material is evaporated and with this also the impurities in the cell as described by Gardner et al. \cite{gardner_Modified_2016}. While Gardner et al. explain the role of the purification of the gallium cell by overheating the cell on the mobility of QWs we show that the optical properties can be enhanced significantly by overheating the aluminum cell.
\end{sloppypar}
\subsection{\label{CV_section}Electrical Characterization}
\begin{figure}[h]
\centering
\includegraphics[width = 8.5cm]{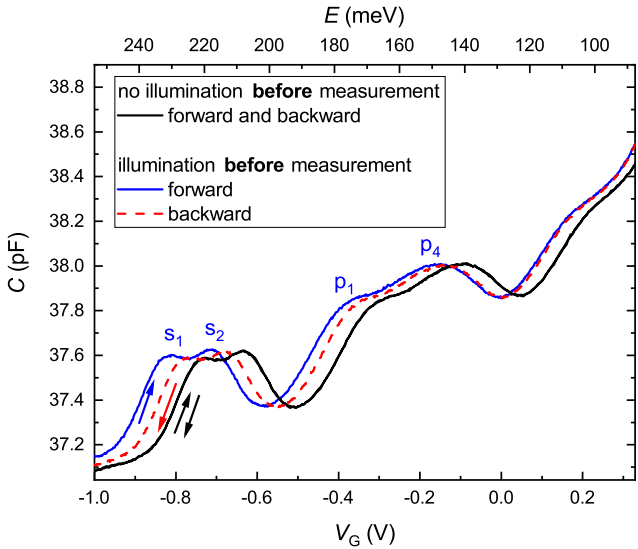}
\caption{Comparison of bidirectional CV measurements of sample $B_{\mathrm{CV,C1}}$ without illumination (black) and with illumination before the measurement using a $920$\,nm IR-LED at $V_{\mathrm{illum}}~=~-2.0$\,V (forward sweep (blue solid), backward sweep (red dashed)). A shift of peak position to more negative $V_{\mathrm{G}}$ and a hysteresis between forward and backward sweep is visible for the measurement with previous illumination.}
\label{comparison_illu_noillu}
\end{figure}

Low-temperature mobility measurements via Hall-effect on van-der-Pauw samples $A_\mathrm{Hall,C2}$ and $B_\mathrm{Hall,C1}$ did not show any deviations to each other or to other samples grown using the reconditioned Al-cell-1$^*$: $1.26\times 10^{6}\,\mathrm{cm}^2/\mathrm{Vs}$ for $A_\mathrm{Hall,C2}$  and $1.27\times 10^{6}\,\mathrm{cm}^2/\mathrm{Vs}$ for $B_\mathrm{Hall,C1}$.\\
However, CV measurements on InAs QDs grown with Al-cell-1 show a large broadening of the full width at half maximum (FWHM) of the peaks in comparison to samples grown with the reconditioned Al-cell-1$^*$ (cf. Fig. \ref{comparison_illu_noillu}).

\begin{figure}[t]
\centering
\includegraphics[width = 8.5cm]{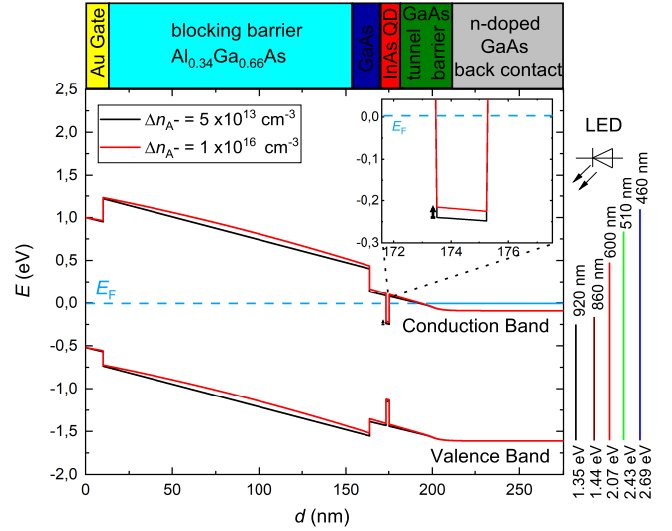}
\caption{Comparison of band structure simulations for an impurity density in the Al region of $\Delta n_{\mathrm{A}^-}~=~5~\times\,10^{13}\,\mathrm{cm}^{-3}$ (black) and $\Delta n_{\mathrm{A}^-}~=~1~\times\,10^{16}\,\mathrm{cm}^{-3}$ (red). Inset shows the enlarged QD region, which was approximated by a QW. Photon energies on the right and sample structure on top are shown for clarity. Shift of the QW energy levels is indicated by a black arrow and the Fermi energy is marked in blue. Schematic of the sample structure on top.}
\label{simulation}
\end{figure}

\noindent In comparison to the CV spectra in Fig. \ref{gute_probe} of a sample grown with the reconditioned Al-cell-1$^*$, only two of four p-peaks are visible. Additionally, we see a shift of the QD CV spectra when the sample is illuminated before the measurement and a hysteresis for different $V_{\mathrm{G}}$ sweeping directions (cf Fig. \ref{comparison_illu_noillu}). Note that the broadening of the peaks is mainly attributed to the inhomogeneity of the QD ensemble and, hence, an increased broadening could also be induced by fluctuations of the QD properties. \\
As we assume the impurity concentration to be homogeneous over the whole aluminum-containing layer, we simulate the energy band structure using an 1D band simulation \cite{snider_1D_} and compare the band structure of the two simulated samples with different ionized acceptor trap state concentrations. One reference sample was simulated with a background impurity density of $\Delta n_{\mathrm{A}^-}~=~5\times 10^{13}\,\mathrm{cm}^{-3}$ in all layers and one with an increased acceptor impurity density of $\Delta n_{\mathrm{A}^-}~=~1~\times10^{16}\,\mathrm{cm}^{-3}$ in the AlGaAs region (see {Fig.~\ref{simulation}}). The QDs in our experiment act as energy probes in CV spectroscopy. We restrict ourselves to a simulation of the conduction band shift at the location of the QDs. 

\begin{figure}[t]
\centering
\includegraphics[width = 8.5cm]{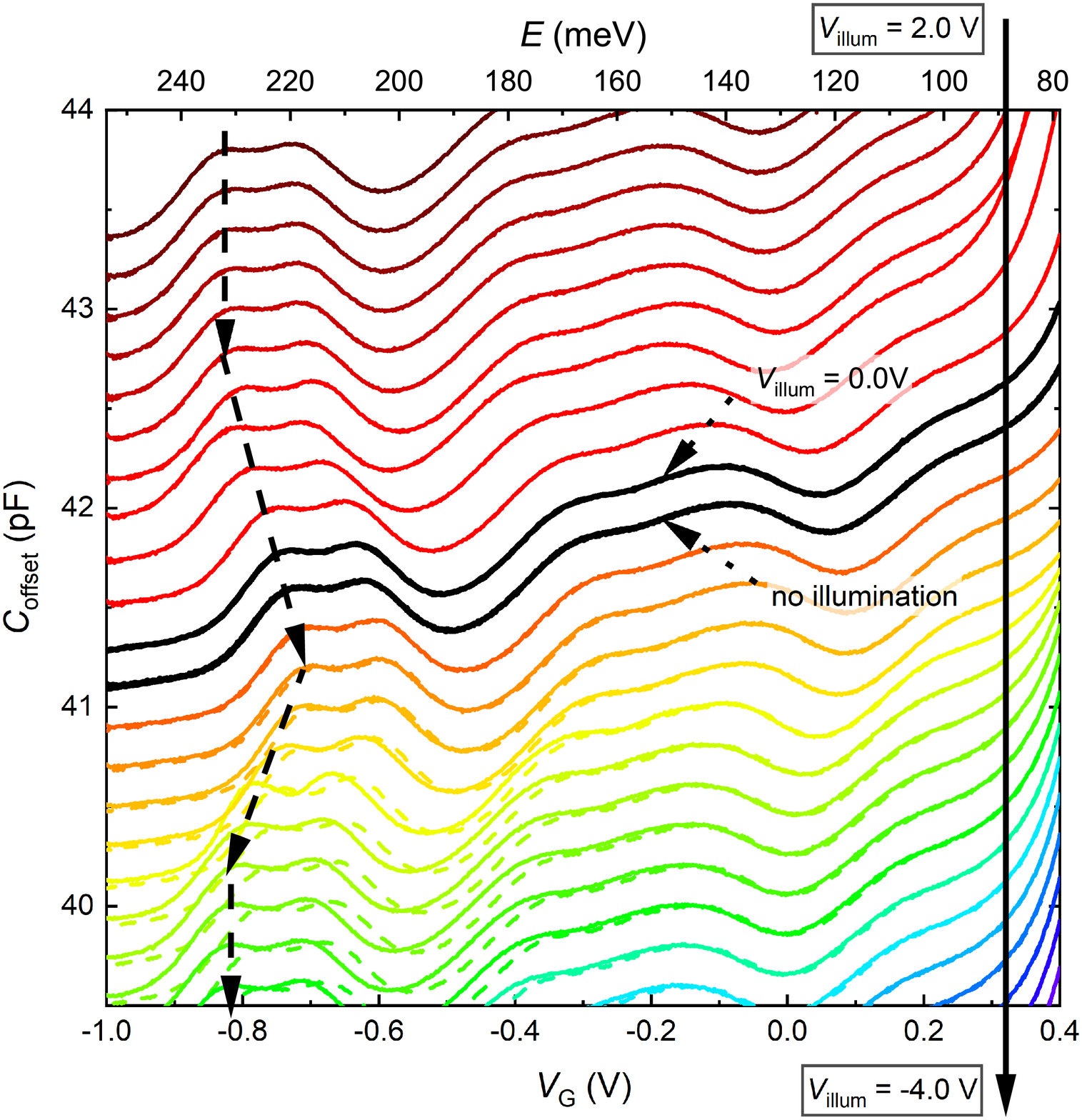}
\caption{Selected parts of waterfall plot of CV illuminated with a $920$\,nm IR-LED at $V_{\mathrm{illum}}~=~2.0\,\mathrm{V}~\mathrm{to}~-4.0\,\mathrm{V}$ in steps of $0.2$\,V before each CV measurement of sample $B_{\mathrm{CV,C1}}$. Forward measurement indicated with solid lines, backward measurement with dashed lines. The measurements are offset with respect to the $V_{\mathrm{illum}} = -4.0\,$V measurement for clarity. No illumination and $V_{\mathrm{illum}} = 0.0$\,V measurements are highlighted black. Dashed arrows show the shift as a guide to the eye. Arrow beyond the diagram indicates further measurements that show no shift.}
\label{14331b_950wf}
\end{figure}

\begin{figure*}[t]
\centering
\includegraphics[width = \textwidth]{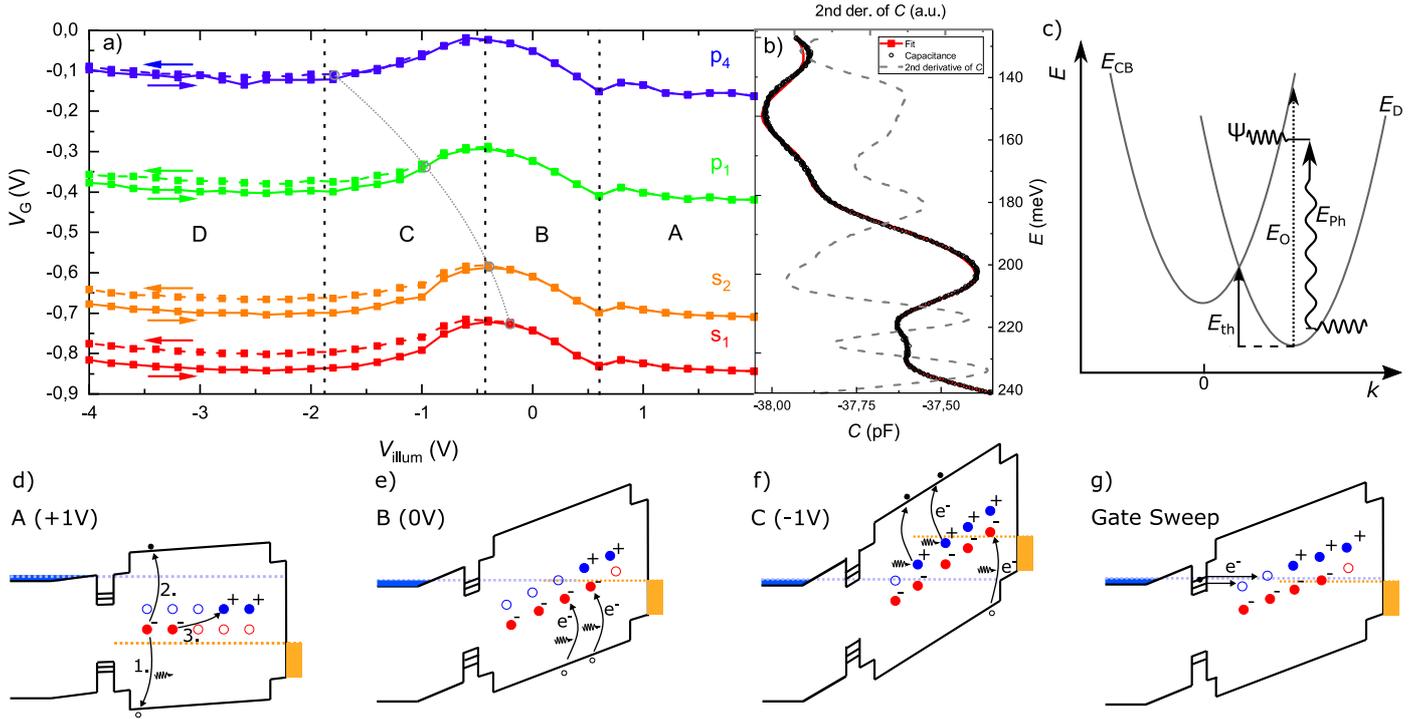}
\caption{a) Peak positions of $\mathrm{s}_1$-, $\mathrm{s}_2$-, $\mathrm{p}_1$- and $\mathrm{p}_4$-charge state using the $920$\,nm IR-LED for $V_{\mathrm{illum}} = 2.0\,\mathrm{V}~\mathrm{to}~-4.0\,\mathrm{V}$ in steps of $0.2$\,V of sample $B_{\mathrm{CV,C1}}$. Arrows show the direction of the sweep. The diagram is divided into four sections (A, B, C, D) for clarity. Grey dotted line as a guide to the eye for the shift of hysteresis opening with higher energy levels. b) CV measurement for $V_{\mathrm{illum}} = 2.0$\,V (dotted), corresponding 2nd derivative for peak position determination (dashed) and cumulative Gaussian Fit (red). c) Illustration of Franz-Keldysh assisted optical ionization of defect states. $E_{\text{CB}}$ shows band diagram of an electron in the conduction band and $E_{\text{D}}$ for an electron in the donor state. For optical ionization $E_{\text{0}}$ of defect states higher energy is required than for thermal ionization $E_{\text{th}}$ due to the shift in $k$-space \cite{mooney_deep_1990}. Due to Franz-Keldysh effect at high voltages optical ionization occurs with photon energy smaller than optical excitation energy ($E_{\text{Ph}}<E_{\text{O}}$) as the wavefunctions $\Psi$ can partly leak into the band gap. d) Schematic of region A: During cool-down donors and acceptors are ionized thermally. With illumination acceptor bound electrons can 1. recombine with illumination induced holes, 2. be excited into the conduction band, 3. recombine with ionized donors. Nearly all acceptors are neutralized in this region leading to saturation in region A of a). e) Schematic of region B: Without illumination this shows thermally ionized defects after cool-down. With illumination further acceptors become persistently ionized (neutralized) if they are below (above) Fermi energy of the gate. f) Schematic of region C: Additionally, donors get ionized due to Franz-Keldysh effect if they are located above the Fermi-level. g) During the gate sweep electrons can tunnel out of the QDs and neutralize donors leading to a hysteresis between forward and backward sweep. }
\label{all_950}
\end{figure*}

\noindent 
With an increased impurity density, an energy band bending is strongly visible in the AlGaAs blocking barrier. Furthermore, the increased acceptor impurity density leads to a shift of the quantum dot energy level of approximately $25$\,meV towards higher energy. Note that simulating the energy band structure with the same trap state density but using donor trap states leads to opposite band bending and a shift of the quantum dot energy level towards lower energy.\\
In CV spectroscopy we expect to see a peak whenever the Fermi-level of the back contact is in resonance with the QD energy level. Accordingly, we expect to see a shift in the CV charging spectrum towards higher (lower) gate voltage if the QD energy level is raised (lowered) with activated acceptor (donor) trap states as a higher (lower) applied field is needed to get the QD into resonance with the Fermi reservoir of the back contact.  \\
We find a shift of charging spectra in the same order of magnitude as predicted from the band simulation for sample $B_{\mathrm{CV,C1}}$ if the sample is illuminated at $V_{\mathrm{illum}}~=~-2.0$\,V using a $920$\,nm IR-LED before the measurement (Fig.~\ref{comparison_illu_noillu}). 
We furthermore find a hysteresis between forward and backward sweeping direction, which is unexpected as the tilting of the energy level in CV is a reversible process and the charging spectrum, therefore, should be independent of the gate sweeping direction \cite{labud_direct_2014}. Thus, the appearance of a hysteresis can be ascribed to different charge environments between forward and backward measurement. Charging peaks in the measurement taken in the forward direction are located at lower gate voltages than in the measurement taken in backward direction. This observation indicates a more positive charge environment in the AlGaAs barrier during the forward sweep.
We further probe the sample with different $V_{\mathrm{illum}}$ at illumination. We see a shift of peak charging voltage first to more positive $V_{\mathrm{G}}$ starting from $V_{\mathrm{illum}} = 0.5$\,V until $V_{\mathrm{illum}} = -0.5$\,V followed by a shift to lower $V_{\mathrm{G}}$ when decreasing $V_{\mathrm{illum}}$ beginning with $V_{\mathrm{illum}} = 2$\,V  ({Fig.~\ref{14331b_950wf}}). This can be understood as a $V_{\mathrm{illum}}$ dependence of the photo-induced activation of trap states.
\noindent
Additionally, the hysteresis is only present for $V_{\mathrm{illum}} < -0.4\,$V.\\
We extract the peak positions of $\mathrm{s}_1$-, $\mathrm{s}_2$-, $\mathrm{p}_1$- and $\mathrm{p}_4$ charge state as a function of $V_{\mathrm{illum}}$ for both sweeping directions ({Fig.~\ref{all_950}} a)) from the 2nd derivative of the CV spectrum. {Fig.~\ref{all_950} b)} shows an exemplary analysis for the CV spectra illuminated at $V_{\mathrm{illum}} = 2.0$\,V (black solid line) before the measurement. The red graph shows a fit to a sum of Gaussians with peak positions extracted from the 2nd derivative (dashed line). 

\noindent The $\mathrm{p}_2$- and $\mathrm{p}_3$- peaks could not be determined reliably due to the above-mentioned broadening and are therefore not included in the analysis. \\
We separate the data into 4 regions: Region A denotes the region at high positive $V_{\mathrm{illum}}$ where no shift in peak position is visible while region B (region C) denotes the region with a shift in peak position to higher (lower) $V_{\mathrm{G}}$. In the last region D, only a slight shift is visible. While the shift in peak position starts in region B for all QD charge states, the hysteresis starts to open at more negative $V_{\mathrm{illum}}$ for higher QD charge states and is less pronounced.\\
We explain the voltage dependence of the photo-induced defect ionization using below band gap excitation originating from a change in optical absorption due to the Franz-Keldysh effect: Via this effect, below band gap absorption is increased as the electron and hole wavefunctions are overlapping into the band gap with increasing electric field (cf. Fig. \ref{all_950} c)) \cite{franz_Einfluss_1958}. Another possible explanation is the $k$-space indirect excitation \cite{mooney_deep_1990} combined with a Franz-Keldysh-like electric field dependent enhancement. Thus, with a higher applied field at illumination more trap states are ionized. \\
In region A, a shift of peak position from no illumination measurement to lower gate voltage is visible (cf. Fig. \ref{14331b_950wf}). This can be explained by the neutralization of thermally ionized acceptors. At high positive $V_{\mathrm{illum}}$ nearly all acceptors are neutralized as the peak position saturates. Fig. \ref{all_950}d) shows a schematic of the possible neutralization processes: Acceptor bound electrons 1.  recombine with illumination induced holes of the valence band, 2. get excited into the conduction band, 3. recombine with ionized donors. Starting in region B we see a shift in peak position to more positive $V_{\mathrm{G}}$ indicating the ionization (neutralization) of acceptor defects, possible due to an increased Franz-Keldysh effect if they are localized below (above) the Fermi-energy of the gate (cf. Fig. \ref{all_950}e)). With increasing electric fields and thus stronger Franz-Keldysh effect also donor trap states get ionized in region C, leading to a reverse shift in peak position (cf. Fig. \ref{all_950}f)). In region D, the shift seems to saturate, as a higher applied field leads to only small changes in peak position. In this region, all energetically available trap states are activated.\\
We explain the hysteresis between forward and backward sweep with the neutralization of activated donor trap states. Without illumination, electrons are the only free carriers in the device. Electrons from the QDs can escape into trap states in the barrier, neutralizing positively charged trap states. This process is illustrated in Fig. \ref{all_950}g). After neutralization, less ionized donor trap states are present during the backward sweep and the energy band bending is increased again due to remaining ionized acceptor traps. Therefore, the peak positions are located at more positive $V_{\mathrm{G}}$ than during the forward sweep. The hysteresis increases with more negative $V_{\mathrm{illum}}$ due to the increased number of trap states activated assisted by the Franz-Keldysh effect.\\
The occurrence of the hysteresis is also proof for the assumption of both acceptor and donor trap states. As the hysteresis effect is explained by the neutralization of positively charged traps during the forward sweep, it is necessary to consider ionized donor traps. Since the hysteresis is not happening at $V_{\mathrm{illum}}$ when the first shift in peak position towards positive voltages in region B occurred, assigned to the activation of acceptor defects, both positively and negatively charged trap states have to be present. \\
The smaller opening of the hysteresis at lower $V_{\mathrm{illum}}$ for higher energetic QD states is explained by the low number of activated donor trap states as most of the neutralization process occurs already at the charging of the s-energy levels. Therefore, a hysteresis for the high QD energy levels is only seen at high negative $V_{\mathrm{illum}}$ when more donor states are ionized due to a larger impact of the Franz-Keldysh effect at high negative $V_{\mathrm{illum}}$. In agreement with the aforementioned explanation of the hysteresis, further repetitions of the $V_{\mathrm{G}}$ sweeps in the same voltage range lead to no new hysteresis, as the same trap states get ionized or neutralized. Also, measurements at different illumination intensities were performed and revealed no intensity dependence.\\
\begin{figure}[t]
\centering
\includegraphics[width = 8.5cm ]{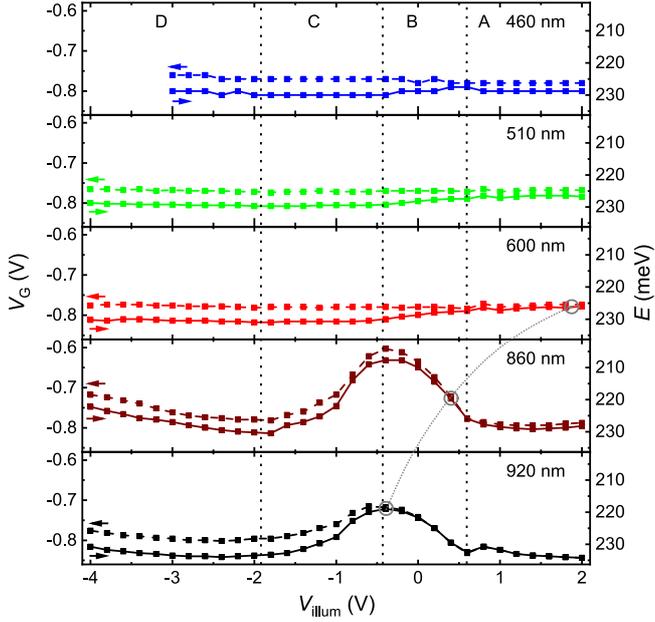}
\caption{$\mathrm{s}_1$ peak position in dependence of $V_{\mathrm{illum}}$ for LED wavelength of $920$\,nm, $860$\,nm, $600$\,nm, $510$\,nm, and $460$\,nm of sample $B_{\mathrm{CV,C1}}$. The analysis is shown for the $\mathrm{s}_1$ charge state only but applies to any other state as they behave the same. Grey dotted line as a guide to the eye for the shift of hysteresis opening with increasing excitation energy.}
\label{s1_comparison}
\end{figure}
\indent However, measurements using different excitation energies revealed an excitation energy dependence of the peak position shift and the hysteresis.
While using below band gap excitation of $860$\,nm and $920\,$nm, a voltage dependence of defect ionization is still visible ({Fig.~\ref{s1_comparison}}), measurements using higher excitation energies ($600$\,nm, $510$\,nm, and $460$\,nm) show no shift in peak position. This can be explained by the immediate ionization of all trap states due to high excitation energy. While the ionization process by IR excitation is assisted by a Franz-Keldysh-like effect, excitation at higher energy leads to a saturation and no shift with decreasing $V_{\mathrm{illum}}$.\\
Increasing the excitation energy leads to an opening of the hysteresis at more positive $V_{\mathrm{illum}}$, i.e. around flat band position, as a smaller Franz-Keldysh effect is sufficient to activate donor trap states \cite{mooney_deep_1990}. While the opening of the hysteresis is still visible using $860$\,nm excitation around $V_{\mathrm{illum}} = +0.4$\,V, a hysteresis is already present at high $V_{\mathrm{illum}}$ for the higher excitation energy LED illumination. These findings support the above mentioned higher ionization energy of donor trap states in comparison to acceptor trap states and the explanation of the hysteresis occurring with the ionization of donor states. The measurement using the $860$\,nm LED shows a small hysteresis at high $V_{\mathrm{illum}}$, which we assume to be a measurement artifact as it vanishes again with decreasing $V_{\mathrm{illum}}$. Note that the ionization of some trap states persist even after temperature ramps between $4$\,K and $300$\,K over a timescale of several days as a shift of the CV spectrum with respect to the non-illuminated spectrum was still seen after five days. \\

\subsection{Model}
In order to quantify the impurity density, one could calculate the charge induced shift analytically \cite{lobl_Narrow_2017}. However, we perform band structure simulations using a 1D Poisson solver \cite{snider_1D_} to calculate the impurity density. We extract the energy shift in the QD region between a structure without and with increased impurity density in the AlGaAs region and find a nearly perfect linear dependence of impurity density $\Delta n_{\mathrm{A}^-}$ and energy shift $\Delta E$ up to $\Delta n_{\mathrm{A}^-}\,=\,2.0\,\times\,10^{16}\,\mathrm{cm}^{-3}$ ($R^2\,=\,0.99994$) (see Appendix Fig. \ref{linreggraph}) given by:

\begin{align}
\Delta n_{\mathrm{A}^-}(\Delta E)  &= a \Delta E
\label{eq_n(E)_lin}
\end{align}
with
\begin{align*}
a &= (-4.149 \pm 0.008) \times 10^{17}\,\frac{\mathrm{cm}^{-3}}{\mathrm{eV}}.
\end{align*}

\begin{figure}[t]
\centering
\includegraphics[width = 8.5cm]{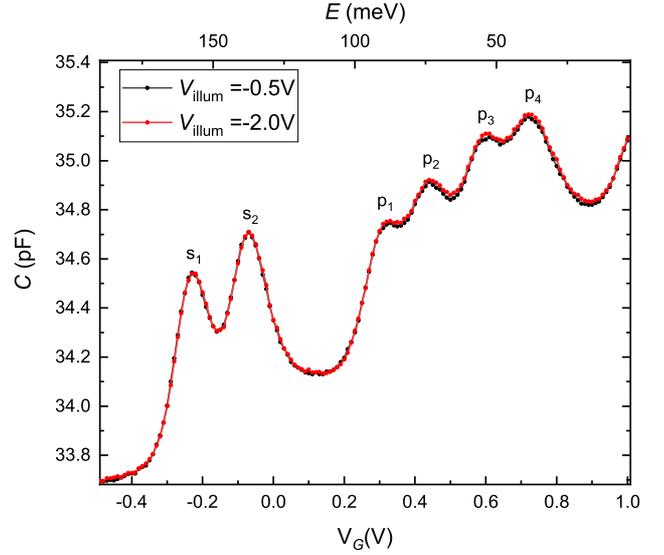}
\caption{CV measurement on sample $A_{\mathrm{CV,C1^*}}$ grown with the reconditioned Al-cell-1$^*$ illuminated with a $920$\,nm LED at $V_{\mathrm{illum}} = -0.5$\,V and $V_{\mathrm{illum}} = 2.0$\,V before the measurement.}
\label{gute_probe}
\end{figure}

\noindent With the maximum energy shift of $25$\,meV between region A and region B using the $860$\,nm LED ({Fig.~\ref{s1_comparison}}) we calculate the impurity density of photo-induced trap charges in the Al region to be approximately $\Delta n_{\mathrm{A}^-}\,=\,(1.04 \pm 0.03)\,\times\,10^{16}\,\mathrm{cm}^{-3}$.\\
\indent To put the the calculated impurity density into comparison, we measured the maximum illumination and electric field-induced energy shift of a reference sample $A_{\mathrm{CV,C1^*}}$ grown with Al-cell-1$^*$ with the same measurement setup and procedure as explained in Sec. III (see {Fig.~\ref{gute_probe}}).
\noindent As the shift in peak position is too small to be extracted in the previous way, we use another method by approximating each peak as Gaussian. The distance between two Gaussians $g_-(V)$ and $g_+(V)$
\begin{align}
g_{\pm}(V) = \frac{1}{\sigma \sqrt{2 \pi}}\mathrm{exp} \left \{-\frac{1}{2}\left( \frac{V \pm \delta V/2}{\sigma}\right)^2\right\} 
\label{gaussians}
\end{align}
shifted by $\delta V$ is calculated by approximating following standardization via Taylor series expansion in first order \cite{ludwig_elektrische_2011}:
\begin{align}
G = \frac{g_1(V) - g_2(V)}{g_1(V) + g_2(V)} \approx - \frac{\delta V}{\sigma^2} V, 
\label{normalization}
\end{align}
where $\sigma = \frac{\mathrm{FWHM}}{2\sqrt{2\ln(2)}}$ denotes the standard deviation. \\
The distance $\delta V$ can be extracted from the slope of the standardization (see Appendix Fig. \ref{guteProbeNormierung}). Applying this method to the $\mathrm{s}_1$ region of the two reference CV spectra,  with $\sigma = 0.06$\,V (extracted from fitting a Gaussian to the peak), we extract a voltage shift of $\delta V = (25 \pm 9) \,\mu$V which corresponds to an energy shift of $\Delta E = (3.3 \pm 1.2)\, \mu \mathrm{eV}$ using the simple lever law. Applying Eq.~\ref{eq_n(E)_lin} and performing another band structure simulation, we relate the energy shift of $3.3\,\mu$eV to an impurity density of photo-induced trap charges of approximately $\Delta n_{\mathrm{A}^-} = (1.4 \pm 0.5) \times 10^{12}\,\mathrm{cm}^{-3}$ illustrating the clear difference to the bad quality sample. Note that we measure photo-induced trap charges, which are only part of the real impurity density but still gives an insight into the purity and high-quality of the second sample.


\section{Conclusion}
We discuss the influence of Al-cell related impurities from a contaminated effusion cell on the optical and electrical properties of QWs and QDs. We show that reconditioning the Al-cell by overheating can lead to enhanced optical properties in QWs similar to those grown with an intact Al-cell. Additionally, we studied the influence of photo-induced activation of trap states unintentionally introduced by a contaminated MBE effusion cell on the QD energy levels. Using CV spectroscopy we measured a shift of the QD energy level depending on the voltage at illumination. We attribute these findings to a photon energy and electric field dependent excitation of donor and acceptor trap states via a Franz-Keldysh-like effect. This energy shift is used to calculate the impurity density of photo-induced trap states with the help of band structure simulation and can be used to estimate the sample quality. In contrast to other methods, we propose a way to quickly and easily determine the heterostructure quality suitable for application on the frequently used QD diode structure of state-of-the-art QD experiments without the need for specific measurements setups, such as a resonant fluorescence setup. This method can be further extended with temperature-dependent measurements which could give hints to the activation energy of the impurity.

\section{Acknowledgments}
\begin{sloppypar}
\tolerance 9999
We gratefully acknowledge support from DFG-
TRR160, DFG project 383065199 LU2051/1-1, BMBF -
Q.Link.X 16KIS0867, the DFH/UFA CDFA-05-06, and the IMPRS-SurMat. G.N.N. received
funding from the European Union's Horizon 2020 Research and innovation Programme under the Marie Sklodowska-
Curie Grant Agreement No. 861097 (QUDOT-TECH).
\end{sloppypar}

\section{Data availability}
The data that supports this work is available from the corresponding author upon reasonable request.

\bibliography{main.bbl}

\appendix*
\section{Additional Figures}

\begin{figure}[h]
\centering
\includegraphics[width = 8.5cm]{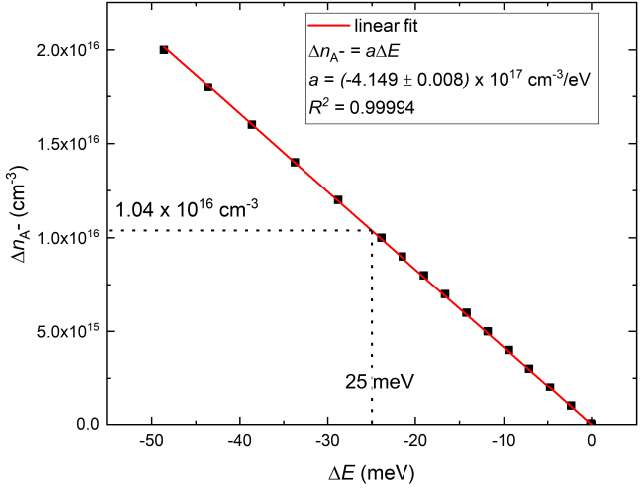}
\caption{Linear regression used to determine Eq.~(\ref{eq_n(E)_lin}). 1D band structure simulations were performed for different impurity densities $\Delta n_{\mathrm{A}^-}$ from which the energy shift $\Delta E$ is extracted.}
\label{linreggraph}
\end{figure}

Fig. \ref{linreggraph} shows the energy shift $\Delta E$ from the band structure simulation for different impurity densities $\Delta n_{\mathrm{A}^-}$ in the AlGaAs region. The dependence is fitted with a linear regression for obtaining Eq. (\ref{eq_n(E)_lin}).\\

\begin{figure}[h]
\centering
\includegraphics[width = 8.5cm]{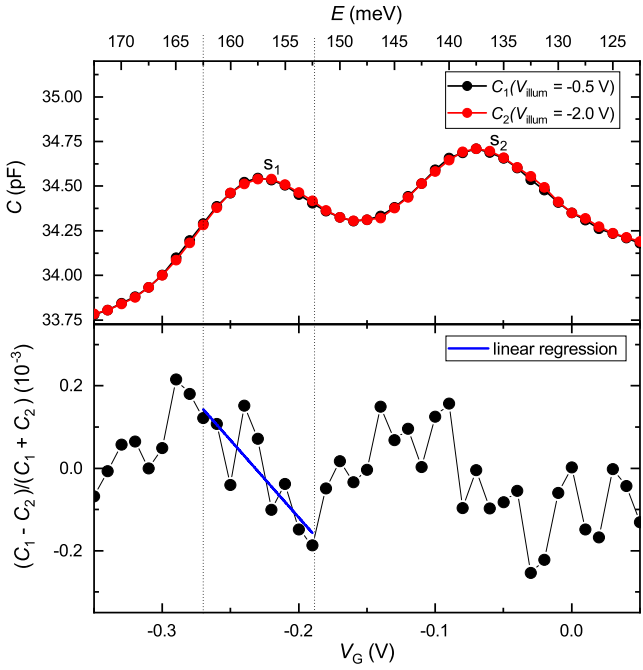}
\caption{a) $s_1$-peak of CV measurements $C_1$(${V_{\mathrm{illum}}~=~-0.5\,V}$) and $C_2$(${V_{\mathrm{illum}}~=~-2.0\,V}$) on the sample $A_{\mathrm{CV,C1^*}}$. b) Standardization of $s_1$-peak after Eq.~(\ref{normalization}) and linear regression used to calculate $\delta V$: $m = (-3.4 \pm 1.2) \times 10^{-3}\, 1/\mathrm{V} \times V_{\mathrm{G}}+ (-7.9 \pm 2.9)\times 10^{-4}$.}
\label{guteProbeNormierung}
\end{figure}

Fig. \ref{guteProbeNormierung} a) shows the s-peaks for a measurement with illumination at $-0.5$\,V and $-2.0$\,V for sample $A_\mathrm{CV,C1^*}$. From the standardization of Eq. (\ref{normalization}) in the $\mathrm{s}_1$-region, the voltage shift $\delta V$ could be extracted with a linear fit. Furthermore, no hysteresis between forward and backward sweep was measured.
\newpage

\end{document}